# Towards optical neuromodulation using nitrogen-doped ultrananocrystalline diamond photoelectrodes

*Samira Falahatdoost[1,#], Andre Chambers[1,#], Alastair Stacey[2], Steven Prawer[1], and Arman Ahnood[1,3,*]*

[1]School of Physics, University of Melbourne, Melbourne, Victoria 3010, Australia

[2]School of Science, RMIT University, Melbourne, Victoria 3000, Australia

[3]School of Engineering, RMIT University, Melbourne, Victoria 3000, Australia

[*] Corresponding author: arman.ahnood@rmit.edu.au

[#]Authors contributed equally

Nitrogen-doped ultrananocrystalline diamond (N-UNCD) is a form of diamond electrode with near-infrared photoresponsivity, making it well suited for physiological applications. N-UNCD's photoresponsivity is strongly influenced by its surface. While it is known that oxygen treatment provides a higher photoresponsivity, a better understanding of its surface processes is needed to tailor the material for optical neuromodulation. This work examines the impact of various oxygen treatment methods, with aim of creating oxygen rich surfaces with different chemical and structural properties. Surface characterisation methods along with electrochemical and photoelectrochemical measurements and modelling were used to investigate the films. It was found that oxygen furnace annealing resulted in orders of magnitude improvement in the near-infrared photoresponsivity, to $3.75 \pm 0.05$ μA/W. This translates to an approximate 200 times increase in the photocurrent compared to the untreated surface. This enhancement in photocurrent is largely due to the changes in the chemical species present at the surface. The photocurrent is estimated to be sufficient for extra-cellular stimulation of brain neurons within the safe optical exposure limit, positioning N-UNCD as an excellent candidate to be used in next-generation photoelectrodes for photobiomodulation.

## 1. Introduction

The field of neural stimulation has a growing number of clinical applications, from treating neurodegenerative disorders such as Parkinson's and Alzheimer's disease [1,2], to implantable bionic devices such as the Cochlear ear implant, bionic eye, and spinal cord stimulator [3–6], Stimulation is conventionally achieved by delivering current from an external power source through conducting electrodes - such as platinum, iridium oxide, titanium nitride, or PEDOT:PSS - in contact with neural tissue [7,8]. However, more recently light-based neural stimulation systems have emerged as an attractive alternative, offering wireless stimulation with superior spatial precision [9]. In particular, light-sensitive electrode materials can be used to transduce light into an electrical stimulus to excite nearby cells [9,10]. In contrast to the widely used light-based technique of optogenetics, this approach to stimulation does not have the ability to differentiate between cell types without the use of targeting agents [11]. However, it avoids the difficulty and ethical implications of genetic modification for the expression of light-sensitive proteins in cells [9].

While there has been much progress in recent years in developing high efficiency photodetectors [12–15], these materials are not necessarily suited for biological environments [9,16]. Of these, silicon-based devices have been the most widely researched in the context of optically-driven neural stimulation, for applications such as retinal prostheses and the long term investigation of neural network dynamics [17–20]. Nevertheless, such devices have as yet limited spatial resolution, while questions remain about the biocompatibility and stability of silicon in solution [21,22].

In consequence, several materials have been investigated as alternative light-sensitive electrodes for neuromodulation. For instance, there has been much work exploring the optoelectronic properties of conductive organic polymers [23–26], inorganic nanomaterials [27] and quantum dots [28][29]. Particularly, organic polymers such as poly(3-hexylthiophene) (P3HT) have recently been shown to exhibit excellent stimulating capabilities for long term optically-driven neural stimulation without significant performance degradation over time [26]. However, for some applications, inorganic semiconductors still possess inherent advantages over soft organic materials, such as the possibility for the fabrication of diverse nano-architectures and the control of signal transduction mechanisms at the interface [30].

In this respect, nitrogen-doped ultrananocrystalline diamond (N-UNCD) is a promising next-generation material for optically-driven neural stimulation, possessing a highly attractive combination of properties such as high chemical inertness, durability, biocompatibility, and large charge injection capacity [31,32]. N-UNCD exhibits semiconducting properties which can be exploited to realize effective electrical and optically-driven stimulation [31,33,34]. We have previously demonstrated that N-UNCD in solution is responsive to light from visible wavelengths to the near-infrared (NIR) [35,36]. Moreover, it displays a capacitive charge transfer mechanism after an oxygen RF-plasma treatment [36], which is preferable to a direct transfer of electrons (Faradaic charge injection). The latter may lead to the formation of harmful reactive oxygen species (ROS) in the solution [7].

Recently, we have demonstrated that the electrochemical capacitance of N-UNCD can be significantly enhanced by furnace annealing in oxygen ambient, thus increasing the charge injection capacity [37]. Surprisingly, the exact method chosen to oxygen terminates the surface has a major impact on the surface capacitance properties. In this paper we report the corresponding increase in the photoresponse and evaluate the performance of N-UNCD as an optically-driven electrode. We investigate the role of different methods of oxygen treatment on the photoresponse of N-UNCD. These methods include (1) furnace annealing in oxygen ambient (OA), (2) oxygen RF-plasma treatment (OP), (3) UV/ozone treatment (UVO), and (4) solution-based oxidation using a mixture of sulphuric acid and sodium nitrate solution (referred to hereafter as 'acid boil' (AB)). We find that, despite the fact that all of these treatments result in an oxygen terminated surface, furnace annealing in oxygen results in the highest capacitive photoresponse by a wide margin. This result is an approximate 100 times enhancement in photoresponse compared to our previous work on oxygen plasma terminated N-UNCD [36]. The physical mechanisms underlying the observed photoresponse of the N-UNCD samples were also investigated in detail through simulation and experimental techniques. These include equivalent circuit modelling based on photocurrent transients, and electrochemical impedance spectroscopy (EIS) to probe the electrochemical properties, as well as Raman spectroscopy and near-edge X-ray absorption fine structure (NEXAFS) spectroscopy to reveal information about the surface composition.

2. Experimental

The preparation process for the N-UNCD films used in this work has been reported elsewhere [37]. Briefly, N-UNCD films were grown on 8 mm × 8 mm pieces of positively charged 4-6 nm nanodiamond seeded crystalline silicon (n+-doped, 1000μm thick, single side polished) using an IPLAS microwave plasma-assisted chemical vapour deposition system in the gas mixture consisted of 79% argon, 20% nitrogen, and 1% methane (*BOC Australia*, 99.999% purity). During the growth, stage temperature was kept at 80 Torr, the microwave power at 1000 W and stage temperature at 850 °C. Our group has previously shown that in spite of nitrogen gas presence during the growth, nitrogen was not detected in the bulk of the diamond grains [38]. The thickness of the N-UNCD films was approximately 32 μm with a diamond grain size previously found to be approximately 5 nm [38]. Acid boiling (AB) in sulphuric acid and sodium nitrate solution, oxygen annealing (OA), oxygen plasma (OP) treatment, and UV/ozone (UV/O) treatment methods were used to terminate the surface of the N-UNCD samples. Acid boil treatment was done in a sulphuric acid/sodium nitrate mixture (3ml of 98.0% *RCI Labscan* sulphuric acid and 0.25 g of 99.0% *chem-supply* sodium nitrate) mixture at ~340°C for 30 minutes in a sand bath, then kept in water for 10 minutes and rinsed several times in DI water. Oxygen annealing treatment was performed in an oxygen environment (gas flow of ~0.2 L/min) in a vacuum furnace at 420°C for 5, 10, 25, and 50 hours. Oxygen plasma treatment was conducted using a *Diener Femto* plasma cleaner, with a mixture of argon: oxygen = 3:1, a power of 50 W, and a pressure of 0.8 mbar for 16 hours in line with previous work [37]. For UV/ozone treatment, the sample was placed in a UV-ozone cleaner (*Bioforce nanoscience UV/Ozone ProCleaner$^{TM}$ Plus,* dominant wavelength: 186/254 nm, intensity: 19.39 mW/cm$^2$) for 20 minutes. H-terminated surfaces were prepared in a 30 Torr 1.2 kW microwave plasma containing hydrogen at 700 °C for 10 minutes.

Raman spectroscopy was used to assess the crystalline quality of samples which has been shown to probe a depth in the order of hundreds of nanometres into the diamond surface [39]. Raman spectra were measured at 532 nm excitation (output power ~ 2.3 mW) using a *Renishaw inVia* spectrometer at room temperature. The spectra were normalized to the maximum intensity and the spline background subtraction was done with *Origin* software. Carbon K-edge Near Edge

X-Ray Absorption Fine Structure (NEXAFS) measurements were conducted at the Soft X-Ray beamline of the Australian Synchrotron. Measurements were conducted in the partial electron yield mode with retarding voltage typically set to 220 V and data was collected from the very near-surface region (<1 nm) of the samples.

Photoelectrochemical measurements were utilized to investigate the photoresponse of N-UNCD samples. These experiments were undertaken using a three-electrode electrochemical cell connected to a potentiostat (*Gamry Interface 1000E*). The cell consists of three electrodes: a platinum disk counter electrode, an Ag/AgCl reference electrode (*eDAQ*), and the N-UNCD sample as a working electrode, as described previously [36]. 0.15 M NaCl solution was used as the electrolyte, chosen as a simple model of physiological conditions. However, it should be noted that real biological environments contain other redox active compounds, while protein fouling of the electrode may also lead to a degradation in performance [40]. A 0.7 W near-infrared (NIR) (808 nm) laser diode (*Wuhan Lilly Electronics*) was utilized as a light source in these experiments, which was pulsed using a waveform generator (*RIGOL DG4062*). The light illuminating the samples had a maximum intensity of 0.41 W mm$^{-2}$ and the illuminated area was 1.7 mm$^2$. This light intensity was chosen to maximise the signal to noise ratio of the photocurrent, while realistic optical parameters for biological applications are considered later in Section 3.4. NIR light has greater optical penetration depth in biological tissue and less phototoxicity than shorter wavelengths [41][42]. The optical setup has been detailed previously [36].

3. Results and Discussion

### 3.1. Photocurrent measurements

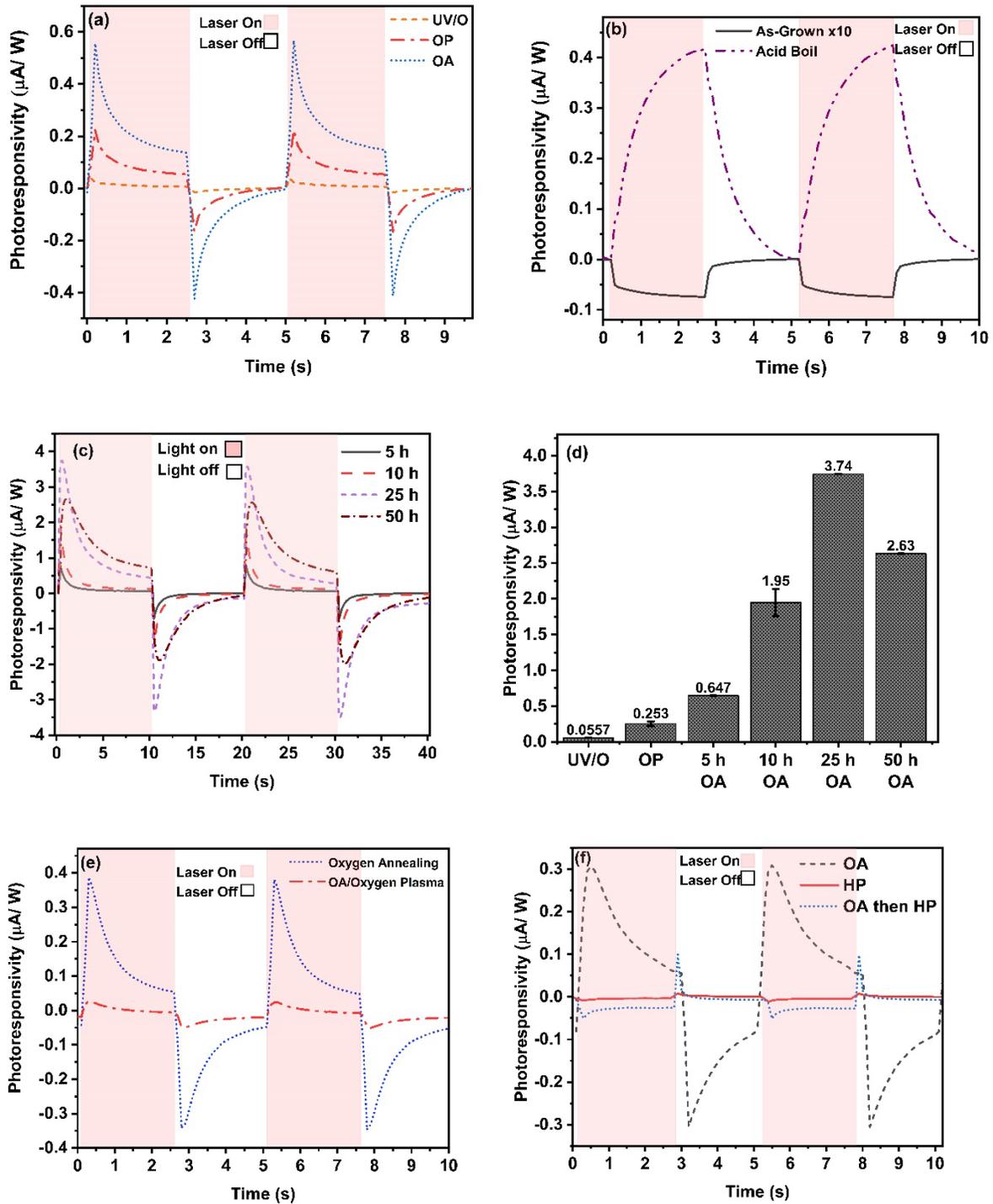

**Figure 1.** Transient photocurrent curves of (a) UV/ozone (UV/O), oxygen plasma (OP), and 5 hours oxygen annealed (OA) N-UNCD samples; (b) as-grown and acid boil treated N-UNCD samples; and (c) N-UNCD samples annealed in oxygen for different times in response to pulsed

illumination at 808 nm with a maximum intensity of 0.41 W mm$^{-2}$ and 1.7 mm$^2$ illuminated area. (d) A histogram of the peak photocurrent values for differently treated N-UNCD samples. (e) shows the photocurrent transient of 5 h-OA N-UNCD compared with the same sample after a subsequent oxygen plasma treatment. (f) Transient photocurrent curves of oxygen annealed and H-plasma (HP) treated N-UNCD samples in comparison with H-plasma post-treatment on the oxygen annealed N-UNCD sample.

As discussed above, we have recently shown that oxygen annealing treatment of the N-UNCD surface dramatically increases the capacitance of the N-UNCD electrode (see **Figure S1** in the Supplementary Information**)** [37]. This finding raises the possibility of using N-UNCD as an optically-driven electrode. We thus measured the photoresponse of the optimised N-UNCD electrode and compared the result to other methods of oxygen termination. The photocurrent of N-UNCD samples was measured while pulsing an 808 nm laser with a maximum intensity of 0.41W mm$^{-2}$ and 1.7 mm$^2$ illuminated area (**Figure 1**). A capacitive transfer of charge in response to light is more desirable compared to a direct (Faradaic) electron transfer, which is known to cause damage to tissue through the introduction of free radicals [31]. During capacitive charge transfer, there is no direct transfer of charge across the photo-illuminated N-UNCD/electrolyte interface, but instead a redistribution of charged chemical species within the electrolyte [7,36]. The potentiostat indirectly measures this capacitive current by the flow of charge through the counter electrode caused by the photogenerated build-up of charge at the working electrode.

The mechanism of charge transfer may depend significantly on two factors: the alignment of the N-UNCD band edges and Fermi level with the energy levels of redox species in solution; and the applied voltage at the interface of the electrode with the electrolyte. It has been shown that for single-crystal diamond, the band edge energy levels differ significantly depending on whether the surface is terminated by oxygen and hydrogen – by about 1.0 eV for single crystal diamond [43]. In particular, the valence band of hydrogen terminated diamond overlaps with the energy distribution of the oxygen redox pair in the electrolyte, facilitating direct charge transfer across the interface [43,44]. On the other hand, the valence band of oxygen terminated diamond does not have a large overlap with the distribution of the redox energy levels, and so Faradaic charge transfer is not facilitated [45]. While N-UNCD has a different structure and composition to single crystal diamond, it is expected that a similar effect would play a role in the observed charge transfer.

Regarding the second factor, it is well known that neural stimulation electrodes should operate within a particular voltage window, exceeding which Faradaic reactions will occur [7]. While the voltage window is not exceeded during light illumination of N-UNCD, it should also be noted that an electrode with larger capacitance results in a lower effective potential at the electrode surface, via the capacitor equation $V=Q/C$. This reduces any risk of approaching the voltage window threshold.

Time-resolved photocurrent measurements provide information about the mechanism of charge transfer at the photoelectrode. A capacitive charge transfer mechanism may be identified by a sharp initial peak in photocurrent followed by a slower decay, with a charge-balanced and biphasic waveform in response to pulsed illumination. On the other hand, Faradaic charge transfer produces a constant photocurrent with a monophasic waveform. Often a combination of the two processes occurs simultaneously, which results in a biphasic waveform that is not charged balanced [46]. Both the capacitance and resistance of the electrode interface affect the observed photocurrent time constant, as is discussed further in Section 3.3.

The results in **Figure 1** show that the method of oxygen terminating the N-UNCD samples has a significant impact on the shape and magnitude of the photocurrent transients. All of the oxygen terminated N-UNCD samples except for the acid boiled sample showed a capacitive charge and discharge in response to light. When the oxygen terminated N-UNCD is illuminated with light, excess free carriers are produced and hole current is generated due to the electric field of C-O dipole moment (**Figure 1(a)**). These photogenerated holes are accumulated at the electrolyte interface and then recombine with electrons from the bulk [36].

As shown in Figure 1(a), the photocurrent of the UV/ozone treated sample ($36.72 \pm 0.85$ nA/W) is negligible compared to other oxygen terminated samples. On the other hand, the photocurrent of the sample that was annealed in oxygen for 5 hours ($0.56 \pm 0.0012$ µA/W) showed the greatest response of all the treatment methods, approximately three times that of the oxygen plasma sample ($0.22 \pm 0.029$ µA/W). These results are in line with our previous measurements of the electrochemical capacitance, where we proposed the enhancements in capacitance were due to a combination of grain boundary etching and oxygen surface functionalities [37]. Surface morphology and roughness as measured by scanning electron microscopy (SEM) and atomic force microscopy (AFM) were found to not be a major factor in the observed results [37]. We then investigated whether the photoresponse could be improved further through a longer annealing treatment. This was done by annealing different samples for

10, 25, and 50 hours (denoted 10 h-OA, 25 h-OA, and 50 h-OA, respectively). Of these, the 25 h-OA sample displayed the greatest photoresponse, with a peak photoresponsivity of $3.75 \pm 0.05$ µA/W which is far greater than previously observed photocurrent in N-UNCD electrodes for neural stimulation [36]. This result was also found to correlate with the electrochemical capacitance as measured by cyclic voltammetry (see **Figure S1** in Supporting Information). The photocurrent transients were also shown to be stable over repeated cycles (see **Figure S3** in Supporting Information).

**Figure 1(b)** shows the photocurrent of as-grown and acid-boiled samples. The as-grown N-UNCD sample has a continuous current flow in response to illumination that is opposite, in polarity, to the oxygen terminated N-UNCD. This behaviour is consistent with a Faradaic charge transfer mechanism that involves the transfer of electrons across the N-UNCD/electrolyte interface to participate in redox reactions in the solution [47]. The Faradaic mechanism of charge transfer can be irreversible and is not favourable for biological applications as it can lead to tissue damage and electrode degradation by forming reactive oxygen species [7]. The surface of the as-grown sample exhibits hydrogen termination after growth and illumination results in the excitation of the near-surface electrons from the valence band and transfer across the interface because of the electric field of the C-H dipole moment.

The acid-boiled sample also shows a monophasic photocurrent waveform suggestive of Faradaic charge transfer but has an opposite polarity to as-grown samples. The opposite polarity of the response can be explained by the opposite direction of the C-O electric dipole compared with the C-H dipole. The photocurrent transient also suggests a high series resistance, which causes the photocurrent to not reduce to zero immediately at the end of the light pulse but rather to gradually decrease. The Faradaic photocurrent of acid boiled N-UNCD may be consistent with surface state mediated charge transfer. Previously, Stacey *et al.* has shown that the acid boil treatment of diamond surfaces introduces $sp^2$ defects on the surface that act as shallow surface states within the bandgap [48]. These surface defects can result in an upward band bending towards the diamond surface and increase the Fermi level in a way that leads to electron transfer through the diamond-electrolyte interface.

The effect of subsequent oxygen plasma treatment of the oxygen annealed sample on the photocurrent of the oxygen annealed N-UNCD sample was also investigated (**Figure 1(d)**). Our previous results have shown that while oxygen annealing removes the trans-polyacetylene (TPA) which is present in the grain boundaries as evidenced by the Raman spectroscopy, subsequent oxygen plasma treatment on oxygen annealed N-UNCD sample results in an increase in the TPA peaks (see **Figure S4** in Supporting Information). This could be explained by the aggressive etching of the N-UNCD surface during the long-term (16 hours) oxygen plasma treatment, which exposes a new surface with TPA content previously inaccessible to Raman spectroscopy. This contrasts with the annealing treatment, which is expected to burn off surface TPA but not etch the diamond grains. It is also noted that thermal diffusion of hydrogen is unlikely to occur during oxygen plasma treatment, which is performed at room temperature. According to Figure 1(d), the oxygen plasma post-treatment on the oxygen annealed sample decreases the capacitive photocurrent in line with capacitance results (Figure S1) [37]. This may indicate that high surface capacitance is negatively correlated with the presence of H and/or TPA on the surface.

This is further supported by the photocurrent results of N-UNCD samples that underwent oxygen annealing and subsequent brief (10-minute) hydrogen plasma treatment as shown in

**Figure 1(f)**. These samples exhibited a dramatic reduction and change in polarity of the photocurrent to resemble the N-UNCD samples that underwent H plasma treatment only, which has similar behaviour to the as-grown sample and has been discussed above This suggests the photoresponse is highly dependent on the chemical species present at the surface, which is consistent with cyclic voltammetry (CV) results of the same surfaces (see **Figure S1(d)**).

### 3.2. Spectroscopic characterization

To better understand the significant increase in photocurrent in the oxygen annealed N-UNCD samples, we have investigated their structural and electronic properties through Raman spectroscopy and NEXAFS spectroscopy. While this analysis has been previously undertaken for UV/ozone, oxygen plasma-treated, acid boiled, and 5 h-OA N-UNCD [37], the effect of longer annealing treatments on N-UNCD has not yet been examined.

**Figure 2** shows the subtracted normalized Raman spectra of as-grown N-UNCD film, 5 h-OA, 10 h-OA, 25 h-OA, and 50 h-OA N-UNCD samples at the excitation wavelength of 523 nm in a range of 500-3500 $cm^{-1}$. A more detailed investigation of the Raman spectra of the as-grown N-UNCD film and oxygen plasma-treated, UV/ozone, treated, oxygen annealed, and acid boiled N-UNCD samples has been discussed in our previous paper [37]. The D-mode of amorphous carbon at 1347 $cm^{-1}$ is the bond stretching of all pairs of $sp^2$ atoms in both rings and chains and the G-mode at 1566 $cm^{-1}$ originates from the breathing mode of $sp^2$ atoms in rings [49]. Our previous Raman results show that oxygen plasma, oxygen annealing, and acid boil treatments of the N-UNCD samples can decrease the ratio of D and G peaks (I(D)/I(G)) and among them, acid boiling decreases the ratio more significantly. Also, a blue shift for D and G peaks was observed for oxygen annealed and acid boiled samples [37].

The Raman spectra in Figure 2 confirm that oxygen annealing treatment decreases the ratio of D and G peaks (I(D)/I(G)), and the D and G peaks also have a blue shift which could be explained by the etching of grain boundaries and surface oxygenation (Table. I). This surface etching of diamond and oxygenation after oxygen annealing has been previously reported [50,51]. Peaks at 1138 cm$^{-1}$ and 1461 cm$^{-1}$ are due to the TPA segment that exist in the grain boundaries [52]. The TPA polymer consists of a long chain of carbon atoms with a repeated unit ($C_2H_2$) that is unstable at high temperatures [52]. We have shown that these peaks are completely removed after acid boiling, while the peaks were flattened after oxygen annealing and to a lesser extent after oxygen plasma treatment [37]. Here, we find that the 5 hours annealing of N-UNCD is sufficient to decrease the intensity of the TPA peaks, while by increasing oxygen annealing time, these peaks are completely eliminated. Because the $v_1$ and $v_3$ modes of TPA are connected to the presence of hydrogen, annealing removes these modes through grain boundary etching [53]. Moreover, thermal oxidation of the TPA polymer through oxygen annealing can remove these peaks and form hydroxyl or per-hydroxyl groups, ketones, and others [54].

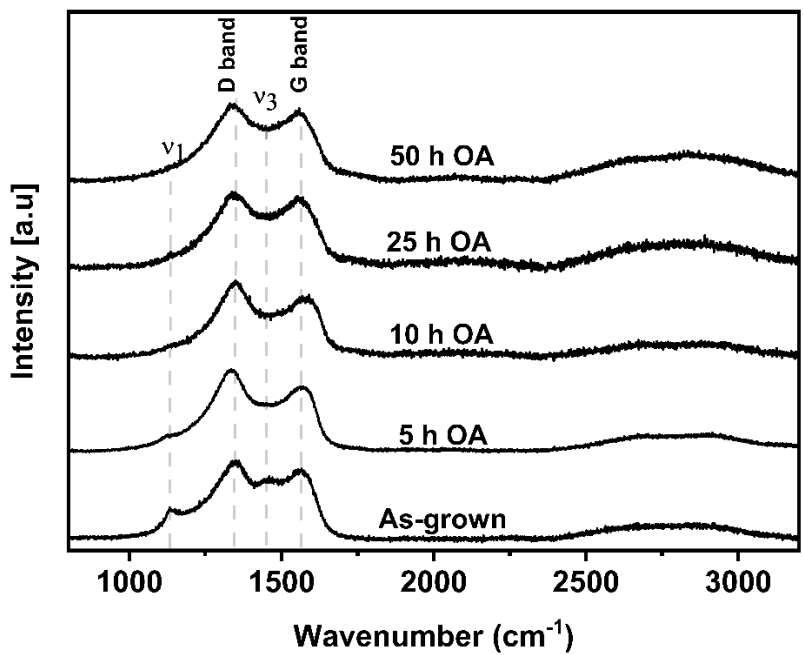

**Figure 2.** Normalized subtracted Raman spectroscopy of as-grown, and N-UNCD film which have undergone oxygen annealed for different times. An excitation wavelength of 523 nm in a range of 500-3500 cm$^{-1}$ was employed.

**Table 1.** D and G peak wavenumbers, I(D)/I(G) ratio, $v_1$ and $v_3$ peak wavenumbers of as-grown and different oxygen annealing time (OA) N-UNCD samples.

| Sample | D (cm$^{-1}$) | G (cm$^{-1}$) | I(D)/I(G) | $v_1$(cm$^{-1}$) | $v_3$(cm$^{-1}$) |
|---|---|---|---|---|---|
| **As-grown** | 1346.9±0.9 | 1566.1±1.0 | 1.15 ± 0.01 | 1138.4±0.9 | 1460.7±1.2 |
| **5 h OA** | 1340.2±0.8 | 1558.3±1.3 | 1.09 ± 0.01 | - | - |
| **10 h OA** | 1337.2±1.0 | 1554.9±1.0 | 1.08 ± 0.01 | - | - |
| **25 h OA** | 1334.9±0.0 | 1553.2±1.0 | 1.06 ± 0.01 | - | - |
| **50 h OA** | 1333.3+0.8 | 1553.3±1.4 | 1.01 ± 0.01 | - | - |

The phase composition of carbon atoms on the N-UNCD surface was also investigated using the C K-edge and oxygen 1s NEXAFS scans of N-UNCD samples with different surface termination that were collected in surface-sensitive total-electron-yield mode (**Figure 3**). The spectra show the expected features of a diamond: the C 1s absorption edge above 289 eV and the second bandgap dip at 302 eV [55]. The peaks located at 285 eV and 291.5 eV correspond to the C 1s-π* transition and 1s-σ* transition of the $sp^2$-bonded carbon graphite, respectively. The π* peak decreases after oxygen annealing treatment, indicating etching of $sp^2$ in the graphitic grain boundary regions. This trend continues after increasing oxygen annealing time to 10 hours, in agreement with the Raman results. The oxidation of the samples is shown by the formation of the peak at 287 eV, originating from the 1s-π* transition (C=O). This peak intensity increases in proportion to the oxygen annealing time.

The bulk and surface NEXAFS spectra of the 10 h-OA N-UNCD sample is shown in **Figure 3(b)**. In bulk mode, NEXAFS can detect species up to a few nanometres deep into the bulk structure [56]. Comparing oxygen annealed N-UNCD bulk and surface, the $sp^2$-bonded carbon (π* peak) decreases in the bulk and almost no oxygen-carbon bond is observed in the oxygen annealed N-UNCD bulk. The oxygen 1s spectrum (**Figure 3(c)**) of the oxygen annealed sample suggests the formation of C-O and C=O groups on the N-UNCD surface after surface termination. This is consistent with previous works showing the formation of C-OH, C=O, and C-O-C bonds at the surface after oxygen annealing [57]. The sharp peak around 531 eV for as-grown N-UNCD may be due to the O-H bond on the surface which has been shown previously [58]. This peak decreases after oxygen surface termination and shows a red shift in line with previous work by Konicek *et al.* [58]. The broad peak starting at 537 eV is from the σ component of single-bonded oxygen [58,59]. The oxygen 1s spectra for the oxygen annealed samples show a notable difference in the C-O peak at approximately 537 eV as compared to other oxygen-treated samples, indicating different surface chemical components on the surface of this sample and could be a factor in the increased electrochemical capacitance and photoresponse. Furthermore, both Raman and NEXAFS results confirm better oxygen coverage on the surface as well as a slight increase in grain boundary etching of the oxygen annealed N-UNCD samples after longer annealing times.

In summary, the Raman and NEXAFS spectroscopy results show that the changes to N-UNCD surface composition after increasing the annealing time are subtle. There is no significant change in $sp^2/sp^2$ carbon content, nor is there much evidence of a change in the $sp^2$ grain size. Despite this the change in capacitance and photoresponse is dramatic. We hypothesise that changes to specific surface chemical functionalities hinted at in the NEXAFS data are the dominant factor responsible for the changes in photocurrent and capacitance. This is supported by the electrochemical measurements of N-UNCD samples that underwent oxygen annealing treatment and a subsequent 10-minute hydrogen plasma treatment, which effectively reversed the effect of the initial treatment. This suggests that etching of the grain boundary regions plays a minimal role in the observed capacitance and photocurrent since the brief H plasma treatment only changed the surface chemical functionalities.

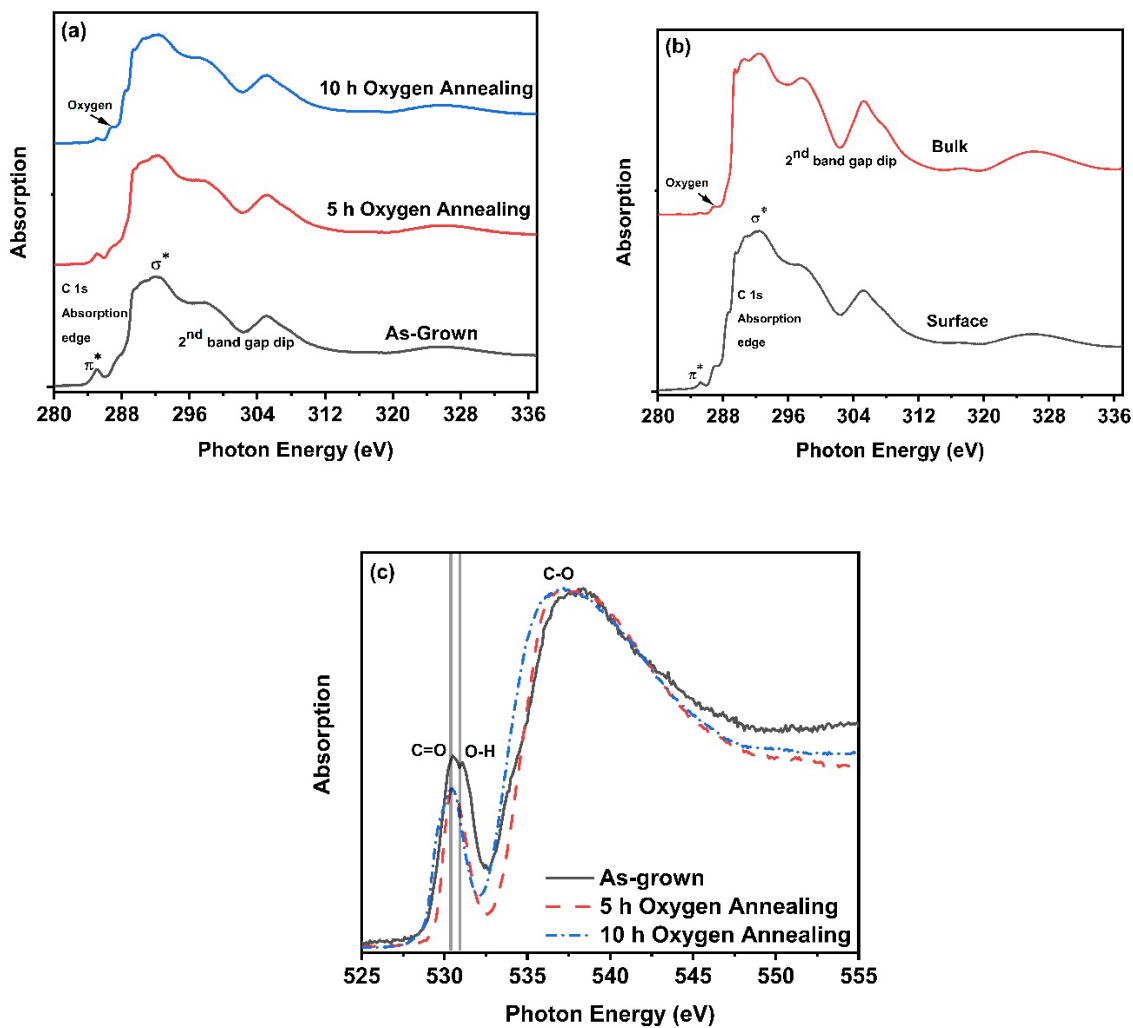

**Figure 3.** NEXAFS spectra for the (a) as-grown, 5 h-OA, and 10 h-OA N-UNCD samples; (b) surface and bulk NEXAFS spectra of oxygen annealed N-UNCD samples; and (c) the oxygen 1s spectra of the surface of the as-grown, 5 h-OA, and 10 h-OA N-UNCD samples.

### 3.3. Photocurrent simulations

In order to integrate the photocurrent results into a larger model of the mechanisms occurring at the diamond/electrolyte interface, simulations of the interface were undertaken using an equivalent circuit model. By fitting the parameters of the model to experimental data, insight can be gained about which properties of the interface are important to the observed electrochemical and photoelectrochemical behaviour.

As shown in **Figure 4**, the model divides the N-UNCD/electrolyte interface into three separate components which correspond to different physical regions. Firstly, the N-UNCD space charge region was modelled as a Schottky diode (governing the current behaviour of the semiconductor-surface state interface) [60,61] in parallel with a constant phase element (accounting for a non-ideal space charge capacitance) [62,63], and a constant current source (accounting for photogenerated current). The electrical double layer (EDL) was modelled using a circuit element governed by the Butler-Volmer equation (accounting for redox current) [60] in parallel with a second constant phase element (accounting for the double-layer capacitance). Finally, the electrical contact and electrolyte were simply modelled as a resistor. For details of this model refer to Supporting Information.

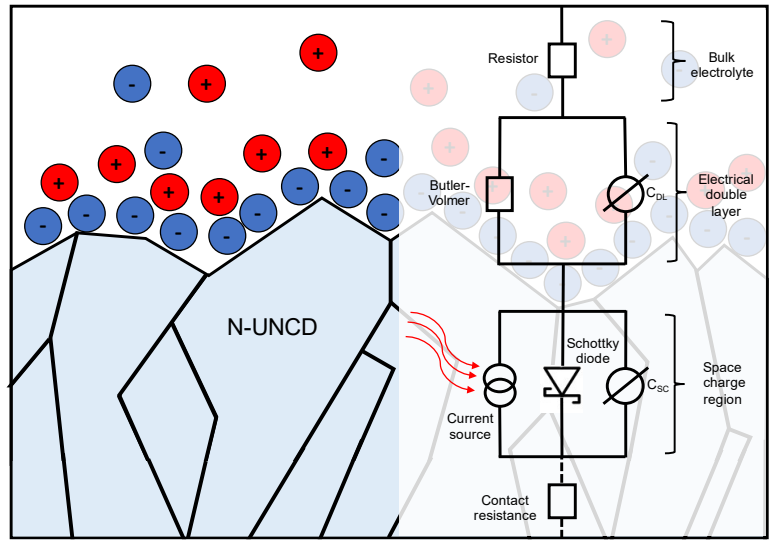

**Figure 4.** The equivalent circuit model of the N-UNCD/electrolyte interface is used for photocurrent simulations. $C_{SC}$ and $C_{DL}$ denote the space charge capacitance and the double layer capacitance, respectively.

The model was fitted to the photocurrent transients shown in Section 3.2, as well as cyclic voltammetry and electrochemical impedance spectroscopy results, with each fitted parameter corresponding to elements in the equivalent circuit (see Supporting Information). As shown in

**Figure 6**, the effective double layer capacitance (C$_{DL}$) and the series resistance (which includes electrolyte and contact resistance) are extracted from the fitted parameters and plotted. The trend in double layer capacitance as a function of surface treatment found by this method is consistent with the capacitance values found by cyclic voltammetry, which is replotted for comparison in **Figure 6(a)**, with some discrepancy expected due to the assumption in CV analysis that the interface behaves like an ideal capacitor.

As discussed above, the photocurrent waveforms of UV/ozone treated, oxygen plasma-treated, and oxygen annealed N-UNCD are indicative of a capacitive charge transfer, with the peak photocurrent correlated with the capacitance of the samples. The simulations confirm that the current flowing through the equivalent circuit is through capacitor charging and discharging. The equivalent circuit model also suggests that longer annealing times correspond with increases in the double layer capacitance, although this effect saturates after 50 hours of annealing time (Figure 6(a)). This enhancement of the capacitance leads to a larger proportion of the photo-excited charge carriers becoming trapped rather than recombining, and so produces a larger total capacitive photocurrent. While the peak photocurrent of the 25 h-OA sample is greater than the 50 h-OA sample, it is noted that the total area under the photocurrent curve (the photo-injected charge) is, in fact, greater for the 50 h-OA sample. This can be explained by the 50 h-OA sample's slightly greater capacitance, but larger series resistance (**Figure 6(b)**) which slows the photocurrent rise time and flattens the curve. This larger resistance is thought to be due to natural variation in contact resistance of the working electrode to the potentiostat.

In contrast, the acid-boiled sample's photocurrent transient displays a gradual increase when illuminated and a gradual decrease when the light is turned off (**Figure 5(d)**). According to the equivalent model fitting, this is due to a combination of high overall circuit resistance but low barrier to charge transfer at the electrolyte interface. As shown in Figure 6(b), the series resistance of the acid boiled sample is approximately five times the resistance of the other samples. When the sample is illuminated, the photogenerated charge carriers flow through the interface by Faradaic charge transfer; however, the high overall circuit resistance significantly slows the photocurrent rise time, and the fall time when the light is turned off. As mentioned above, it is thought that this Faradaic charge transfer is mediated by surface states such as primal $sp^2$ defects found previously in acid-boiled single-crystal diamond [48]. When the light is turned off, the charge stored in the space charge region discharges, leading to a decreasing positive current.

In the case of the as-grown sample, the charge transfer is suggestive of Faradaic processes, but with the opposite polarity of the acid boiled sample, which is thought to be due to the opposite polarity of the C-H dipole at the surface in comparison with the C-O dipole [64,65]. The Faradaic nature of the charge transfer we suggest is due to the proximity of the valence band relative to the energy distribution of redox ions in solution, as discussed in Section 3.1.

The equivalent circuit model fitting reveals that the properties of the electrical double layer region (see Figure 4) are critical to the observed electrochemical behaviour. In particular, the double layer capacitance is central to the changes in CV area and photocurrent transient magnitude. This property is not obvious from Raman and NEXAFS spectroscopy, which only showed subtle changes in the surface composition due to different oxygen surface treatments. This highlights the sensitivity of CV and photocurrent measurement techniques to changes in chemical surface termination.

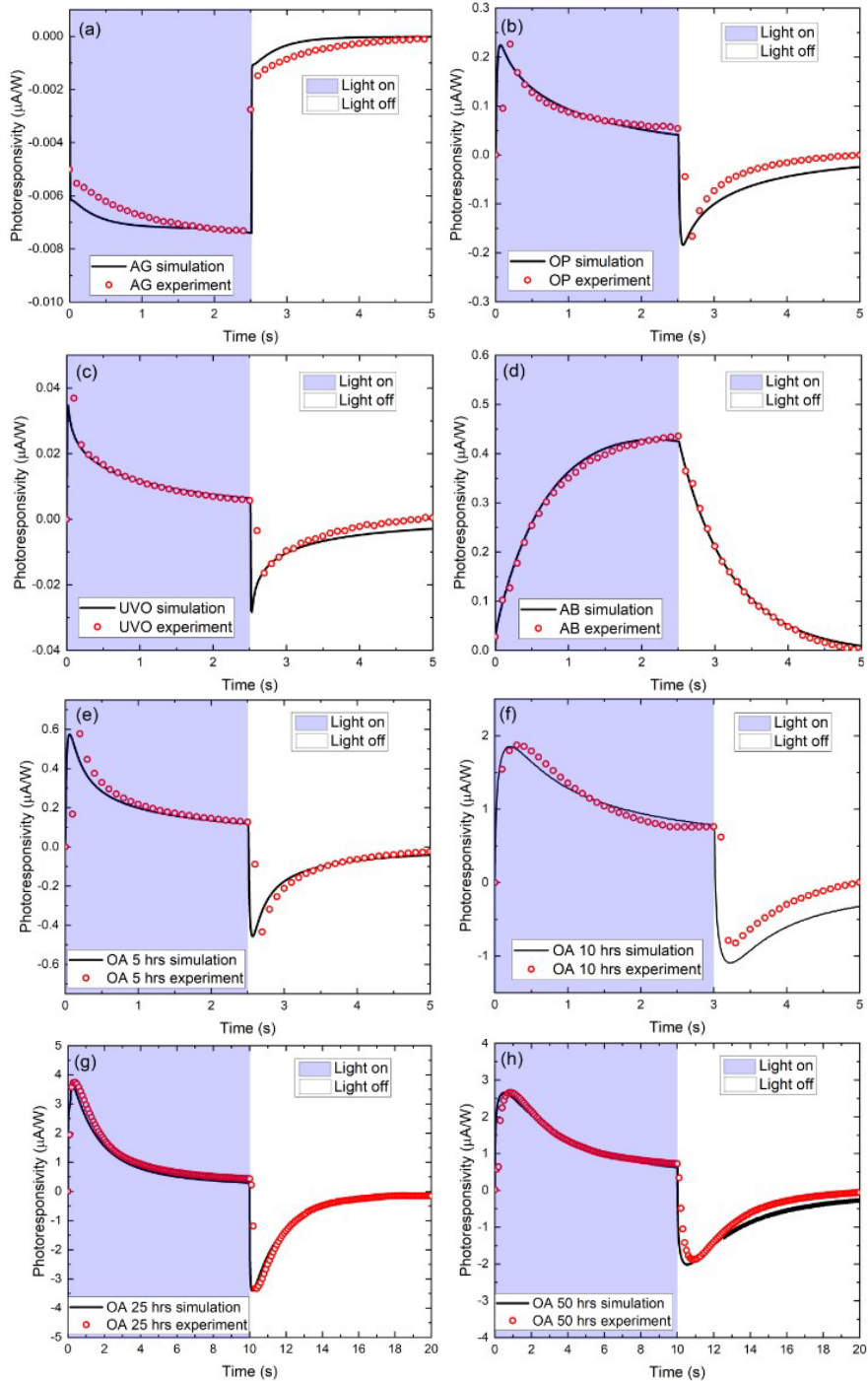

**Figure 5.** Photocurrent simulations of (a) As-grown (AG), (b) oxygen plasma-treated (OP), (c) UV/ozone treated (UVO), (d) acid boiled (AB), and oxygen annealed (OA) for (e) 5 hours, (f) 10 hours, (g) 25 hours, and (h) 50 hours N-UNCD samples in response to pulsed illumination at 808 nm with a maximum intensity of 9.55 W mm$^{-2}$ and 0.26 mm$^2$ illuminated area. The light is turned

on at 0 seconds and turned off at 2.5 seconds for (a-f) and 10 seconds for (g-h). Note the change of vertical scale between plots.

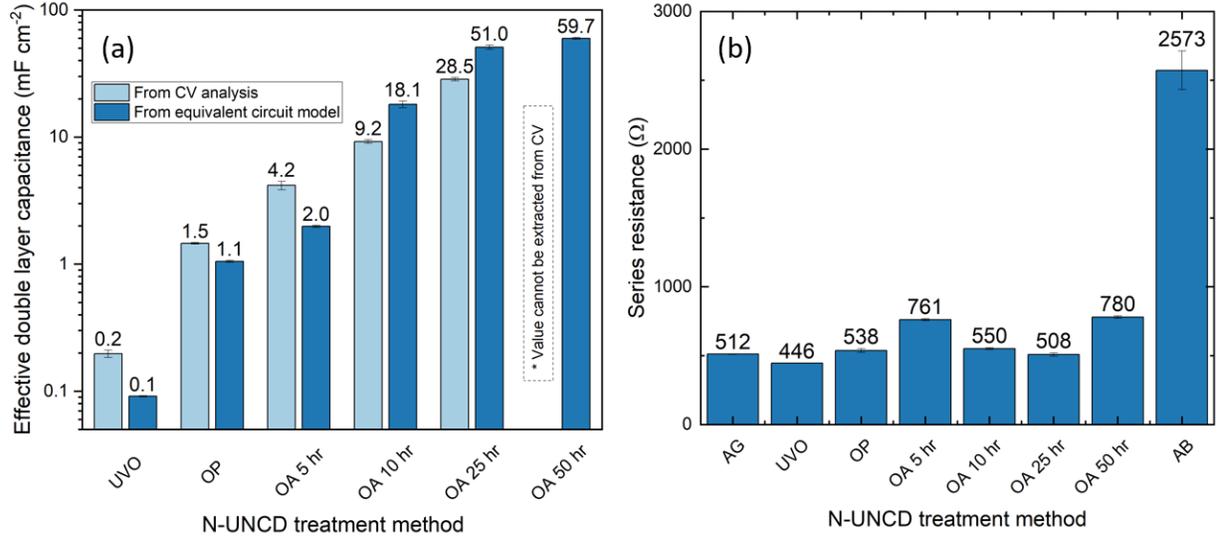

**Figure 6.** Histograms of (a) effective double layer capacitance, and (b) series resistance from the photocurrent simulations.

### 3.4. Evaluation of the N-UNCD electrode for optically-driven neural stimulation applications

To determine the suitability of the optimised N-UNCD electrodes for neural stimulation, the photocurrent results should be compared to known stimulation thresholds. Comparing N-UNCD to the current state-of-the-art interfaces for neural photostimulation, the photoresponse is at least two orders of magnitude lower [24,27]. However, a key advantage of N-UNCD over soft polymer interfaces is the ability to produce nano-architectures and control the mechanism of charge injection [36,66]. Furthermore, conventional inorganic semiconductors such as silicon often display limited biocompatibility [21]. This places N-UNCD as a more favourable material for fabricating multi-electrode arrays and investigating different excitatory processes in neural networks. Therefore, a theoretical evaluation of stimulation efficacy was carried out.

**Table 2.** A selection of previous works on photoelectrodes for neural stimulation.

| Work | Material | Wavelength (nm) | Light responsivity (µA/W) | Successful cell stimulation |
|---|---|---|---|---|
| **This work** | N-UNCD | 808 | 3.75 | - |
| **Karatum *et al.* (2021)** [67] | InP/ZnO/ZnS : PCBM | 445 | ~ 1000 | - |
| **Chambers *et al.* (2020)** [36] | N-UNCD | 808 | 0.0371 | - |
| **Ferlauto *et al.* (2018)** [68] | PEDOT:PSS bottom anode, P3HT:PCBM semiconductor layer, Ti top cathode | 565 | ~ 2100 | Yes (*in vitro*) |
| **Rand *et al.* (2018)** [69] | Cr/Au : $H_2Pc$ : PTCDI | 660 | ~ 6500 | Yes (*in vitro*) |
| **Bareket *et al.* (2014)** [27] | Semiconductor nanorod : carbon nanotube film | 405 | ~ 350 | Yes (*in vitro*) |
| **Ghezzi *et al.* (2011)** [24] | rr-P3HT:PCBM | 530 | ~ 270,000 | Yes (*in vitro*) |
| **Starovoytov *et al.* (2005)** [70] | p-type Si, -0.6 V applied bias | 660 | ~ 300,000 | Yes (*in vitro*) |

Multiple factors impact the effectiveness of an electrode for optically-driven neural stimulation, including the charge injection density, light intensity, proximity to neural tissue, and the type of tissue being stimulated [35]. However, two main factors for consideration in this analysis are the charge injection density, which must be sufficient for extra-cellular stimulation, and the light intensity, which must be within the safe optical exposure limits. These parameters place limits on the minimum illumination time required before sufficient charge is injected to be able to stimulate a neuron. For the laser spot size used in this study, the safe optical exposure limit in Watts may be calculated using the formula $4.62 \times 10^{-2} \times 10^{0.002(\lambda-700)} t^{-0.25}$, where $\lambda$ is the wavelength of light in nanometres, and $t$ is the pulse duration in seconds [71]. Considering the wavelength used in this study (808 nm) and assuming a linear relationship between light intensity and peak photocurrent, the charge injection density of the optimised N-UNCD electrode within the safe optical exposure limit was calculated as shown in **Figure 7**. This was done by integrating the area under the photocurrent transient curves and dividing by the laser spot area. The results indicate that a light pulse duration between 200 ms and 400 ms would be required to meet the threshold range of 2.3-6.7 $\mu C\ cm^{-2}$ for extra-cellular stimulation of brain neurons [72], but the threshold of 23 $\mu C\ cm^{-2}$ for stimulation of the retina is likely too high for any pulse duration [73]. This suggests that N-UNCD may not be suitable for high-frequency stimulation of neurons, but could still be useful for the long-term promotion of neural growth, where the stability and biocompatibility of the material provide distinct advantages. In particular, the area of photobiomodulation has attracted a great deal of interest as a potential therapy for the management of chronic pain, depressive disorders and tissue regeneration [74,75]. N-UNCD may prove to be a very useful material to increase the efficacies of such treatments.

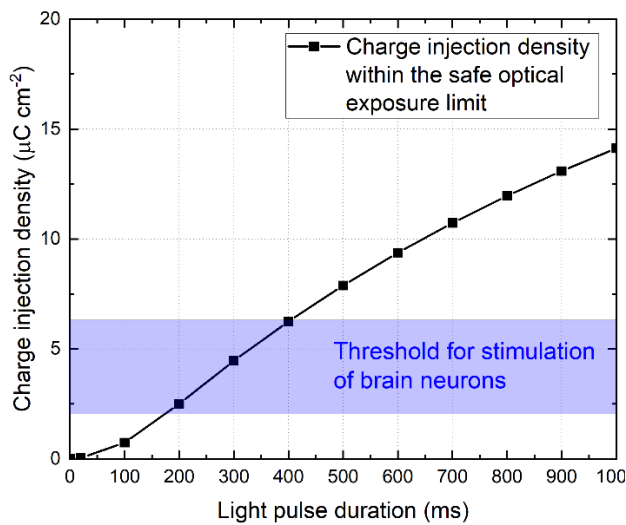

**Figure 7.** The plot of the charge injection density of the 25 h-OA N-UNCD within the safe optical exposure limit, showing the threshold for the stimulation of brain neurons reported by Kuncel *et al.* [72].

## 4. Conclusion

We have demonstrated that different methods of oxygen surface treatments have a major impact on the photosensitivity of N-UNCD. Of these techniques, a simple oxygen furnace annealing procedure produced the optimal interface for the use of N-UNCD as a photoelectrode for neuromodulation with the greatest NIR light responsivity of $3.75 \pm 0.05$ µA/W, which is an approximate 100 times improvement compared to previously reported N-UNCD photoelectrodes. Through equivalent circuit modelling of electrochemical data, we determined that the sub-bandgap photoresponse of N-UNCD electrodes largely correlates with the electrochemical capacitance produced by each treatment method. While previous work has only previously offered preliminary explanations for this enhancement, we suggest these changes are determined mainly by the modification of surface chemical functionalities, not grain boundary etching or thermal diffusion of hydrogen. Moreover, while the photoresponse of the optimised N-UNCD does not exceed other photoelectrode materials of previous works, the electrode can nevertheless generate sufficient photocurrent for extracellular stimulation of brain neurons within the safe optical exposure limit. Indeed, N-UNCD possesses inherent advantages such as the ability to fabricate diverse nanostructures and to control the mode of charge injection. Taken together, this suggests great promise for the future use of N-UNCD as a photoelectrode for long-term photobiomodulation applications, such as tissue regeneration and the treatment of chronic pain.

**Supporting Information**
Supporting Information is available from the Wiley Online Library or the author.

**Declaration of competing interest**

SP is a shareholder in iBIONICS, a company developing a diamond based retinal implant. SP is a director and shareholder of Carbon Cybernetics, a company developing a carbon fibre based neural implant.

**Acknowledgments**

AS is supported by the Australian Research Council (ARC) via fellowship (DE190100336). AC is supported by an Australian Government Research Training Program (RTP) Scholarship. The NEXAFS research was undertaken on the Soft X-ray Spectroscopy beamline at the Australian Synchrotron, part of ANSTO.